\documentclass[twocolumn]{aastex631}
\shorttitle{Dissecting the $\gamma$-ray emissions of NGC 1068 and NGC~253}
\shortauthors{Ji et al.}

\newcommand{\gr}{$\gamma$-ray}

\newcommand{\fermi}{{\it Fermi}}
\newcommand{\ngc}{NGC~1068}

\begin{document}

\title{Dissecting the $\gamma$-ray emissions of the nearby galaxies NGC 1068 and
NGC 253}

%%\correspondingauthor{Zhongxiang Wang}
%%\email{wangzx20@ynu.edu.cn}

\author{Shunhao Ji}
\affiliation{Department of Astronomy, School of Physics and Astronomy, Yunnan
University, Kunming 650091, China; wangzx20@ynu.edu.cn}

\author[0000-0003-1984-3852]{Zhongxiang Wang}
\affiliation{Department of Astronomy, School of Physics and Astronomy, Yunnan
University, Kunming 650091, China; wangzx20@ynu.edu.cn}
\affiliation{Shanghai Astronomical Observatory, Chinese Academy of Sciences, 80
Nandan Road, Shanghai 200030, China}

\author{Yi Xing}
\affiliation{Shanghai Astronomical Observatory, Chinese Academy of Sciences, 80
Nandan Road, Shanghai 200030, China}

\author{Dahai Yan}
\affiliation{Department of Astronomy, School of Physics and Astronomy, Yunnan
University, Kunming 650091, China; wangzx20@ynu.edu.cn}

\author{Jintao Zheng}
\affiliation{Department of Astronomy, School of Physics and Astronomy, Yunnan
University, Kunming 650091, China; wangzx20@ynu.edu.cn}

\begin{abstract}
Intrigued by recent high-energy study results for nearby galaxies with 
\gr\ emission and in particular NGC~1068 that has been detected as a 
neutrino-emitting source by the IceCube Neutrino Observatory, 
we conduct detailed analysis of the $\gamma$-ray data for the galaxies NGC~1068
	and NGC~253, obtained with the Large
	Area Telescope onboard {\it the Fermi Gamma-ray Space Telescope}.
	By checking for their possible spectral features and then constructing 
	light curves in corresponding energy ranges, we identify 
	spectral-change activity from NGC ~1068 in $\geq$2\,GeV energy range
	and long-term detection significance changes for NGC~253 
	in $\geq$5\,GeV energy
	range. In the former, the emission appears harder in two half-year
	time periods than that in the otherwise `quiescent' state.
	In the latter, a $\sim$2-times detection significance
	decrease after MJD~57023 is clearly revealed by the test-statistic
	maps we obtain.
	Considering studies carried out and models proposed for 
	the $\gamma$-ray emissions of the two sources, we discuss 
	the implications of our findings. We suspect that the jet 
	(or outflow) in NGC~1068 might contribute
	to the \gr\ emission. The nature of the long-term detection
	significance change for
	NGC~253 is not clear, but since the part of the GeV emission
	may be connected to the very-high-energy (VHE) emission from 
	the center of the galaxy, it could be further probed
	with VHE observations.
\end{abstract}


\section{Introduction} 
\label{sec:intro}

The launch of {\it the Fermi Gamma-ray Space Telescope (Fermi)} and the
observations carried out with the large Area Telescope (LAT) onboard it
have confirmed the theoretical expectations (e.g., \citealt{vkw89,pag+96,dt05})
that star-forming galaxies can emit significant $\gamma$-rays \citep{abd+10a}.
Due to high density of cosmic rays, emanated from copious supernova 
remnants that are related to the strong star formation 
(e.g., \citealt{ack+12} and references therein), such \gr\ emission 
is produced through the proton-proton collisions and/or leptonic processes 
(i.e., bremsstrahlung or inverse Compton scattering; e.g., \citealt{der86}).
Thus far, more than 10 star-forming galaxies, within the local group or
nearby, have been detected at $\gamma$-rays (e.g., \citealt{aje+20,xi+20,xw23}).
These galaxies lie along a correlation line of the \gr\ versus the infrared
(or radio) luminosities \citep{abd+10b,ack+12,aje+20,xi+20}, which is
considered to
indicate the high cosmic-ray densities related to their star-formation property.

However along with this understanding, there are complications. Some of 
the galaxies
contain an active nuclear or other related components (e.g., the jet or 
outflow), which can contribute \gr\ emissions. The extreme cases are NGC~3424
and UGC~11041, which made themselves detectable with the \fermi\ LAT by having
flare-like events \citep{pen+19}. Another interesting case is \ngc, 
as it has been shown that its \gr\ emission is more intense than
that expected from modeling of the starburst and related cosmic-ray density
\citep{len+10,yoa+14,eb16,mak23}. Alternative sources that possibly power 
the \gr\ emission have been proposed, such as the jet \citep{len+10} or the 
outflow \citep{lam+16} in it. Recently,
very-high-energy (VHE) TeV neutrino emission has been detected from
this galaxy by the IceCube Neutrino Observatory (IceCube; \citealt{ice22}),
making it the second known extra-galactic neutrino-emitting source. 
By taking the neutrino emission into consideration, \citet{eic+22} have 
considered a combination of
two zones, the starburst region plus the nuclear corona, 
and \citet{inos+22} have suggested the disk winds as the power source. 

%Theoretical studies point out that the neutrino emission likely arises from the nuclear region, or more specifically the corona surrounding the supermassive black hole \citep{mkm20,ino+20}.  The processes that produce the neutrinos involve hadronic interactions, while leptonic processes should also occur as the accompanying ones. High or VHE \gr\ emission is produced as well in the hadronic and leptonic processes, but because the optical depth of the corona to the \gr\ photons is high (e.g., \citealt{mur22}), the photons can not escape from the region until they cascade to MeV energy photons. 

To fully understand the physical properties of \ngc\ and other galaxies of
the same group (e.g., \citealt{ack+12}), we have conducted detailed studies 
of their \gr\ emissions, in which we focused on the variability analysis. 
In this paper, we report our analysis results for \ngc\ and another nearby
star-burst galaxy NGC~253. Spectral and/or flux variations were found,
which add more features for helping our understanding of possible physical 
processes 
occurring in them.  Below we first briefly summarize the properties of the two 
galaxies related to this work
in Section~\ref{sec:src}.  The analysis and results are presented in
Section~\ref{sec:ana}.  In Section~\ref{sec:dis}, we discuss the results.

\subsection{Properties of \ngc\ and NGC~253}
\label{sec:src}

At a distance of 10.1$\pm1.8$\,Mpc \citep{tul+08},
%%$\sim$14.4\,Mpc \citep{mey+04}, 
\ngc\ is a nearby galaxy that 
has been extensively studied at multi-wavelengths (e.g., \citealt{pas+19,ino+20}
and references therein). It contains an active nucleus in the center, viewed by
us with an angle of $\sim$70\arcdeg\ \citep{lop+18} and thus classified as type
2. Its GeV \gr\ emission was detected in early observations with the \fermi\ 
LAT \citep{len+10,ack+12}, and no $>$200\,GeV very-high-energy (VHE) emission 
has been detected
\citep{acc+19}. Given it being an active galactic nucleus (AGN), its GeV \gr\ 
emission was searched for variability in studies 
\citep{len+10,ack+12,aje+20,amm23}, but no significant variability was found.

NGC 253 is one of the nearest galaxies to us (at a distance of 
3.5$\pm$0.2\,Mpc; \citealt{rek+05}) and has also been extensively studied at 
multi-wavelengths. It is nearly edge-on with an inclination angle of
$i\sim 72\arcdeg$ (e.g., \citealt{puc+91}) and appears with an approximate 
angular size of 0\fdg48$\times$0\fdg10 in the optical \citep{pat+03}.
One particular feature of it is its starburst in the nuclear
region (e.g., \citealt{rlw88,ott+05}). Related to the starburst, 
VHE and GeV \gr\ emission from it has been detected with the ground-based 
High Energy Stereoscopic System (HESS; \citealt{ace+09})
and \fermi\ LAT \citep{abd+10a}, respectively.
Whether this galaxy contains an AGN or not is under investigation
(e.g., \citealt{wea+02},\citealt{mul+10}, and \citealt{grv20} and
references therein).
%%\citealt{abr+12}, \citealt{per+18}, and references therein). 

\section{Data Analysis and Results} 
\label{sec:ana}

\subsection{Fermi-LAT Data and Source Model} 
\label{ssec:data}

We selected 0.1--500 GeV LAT events (evclass=128 and evtype=3) from the 
updated Fermi Pass 8 database in a time range of from 2008-08-04 15:43:36 (UTC)
to 2023-04-06 00:00:00 (UTC), approximately 14.7\,yr. The region of interest 
(RoI) for each target was set to be
20\arcdeg $\times$ 20\arcdeg\ centered at the center of the galaxy.
For NGC~1068, R.A. = 40\fdg6696 and Decl. = $-0\fdg01329$ (equinox J2000.0),
and for NGC~253, R.A. = 11\fdg8881 and Decl. = $-25\fdg2888$ (equinox J2000.0).
We excluded the events with zenith 
angles greater than 90\degr\ to reduce the contamination from the Earth limb 
and included those with good time intervals (selected with the expression 
DATA$\_$QUAL $>$ 0 \&\& LAT$\_$CONFIG $=$ 1). The instrumental response 
function P8R3$\_$SOURCE$\_$V3 and the software package of Fermitools--2.2.0 
were used in the analysis.

Source models were generated based on the Fermi LAT 12-year source catalog 
(4FGL-DR3; \citealt{4fgl-dr3}) by running the script make4FGLxml.py.
They each included all sources in 4FGL-DR3 within a radius of 25\degr\ 
centered at each of the targets. The positions and the spectral parameters 
of the sources are provided in the catalog. 
In our analysis, the spectral parameters of the sources within 5\arcdeg\ from 
a target were set free and the other parameters were fixed at their catalog 
values. The spectral model gll$\_$iem$\_$v07.fit was used for the Galactic 
diffuse emission, and the spectral file iso$\_$P8R3$\_$SOURCE$\_$V3$\_$v1.txt 
for the extragalactic diffuse emission. The normalizations of these two 
diffuse components were always set free in our following 
analyses.

\begin{table}
	\centering
        \caption{Likelihood analysis results for NGC 1068}\label{tab:ts}
        \begin{tabular}{lccc}
        \hline
		Data/Model & $F_{0.1-500}$/10$^{-9} $  & $\Gamma$  & TS  \\
                  & (photons s$^{-1}$ cm$^{-2}$)  &  &  \\ \hline
		Catalog values   & ---  & 2.34$\pm$0.05 & 337 \\
		Whole data       & 10.2$\pm$1.1   & $2.36\pm 0.06$ & 402  \\
		%%2--500\,Gev  & 1.79$\pm$0.24  & 2.26$\pm$0.15 & 166\\
		P$_q$  & 10.6$\pm$1.2    & 2.43$\pm$0.06 & 320 \\
		P$_1$  & 9.1$\pm$3.8   & 2.02$\pm$0.18 & 52 \\
		P$_2$ &  7.1$\pm$2.8  &  1.85$\pm$0.17 & 47\\
	\hline
\end{tabular}
	\end{table}

For each of the two targets, we set baseline spectral models following those
given in the catalog. The \gr\ counterpart to NGC~1068 was modeled as a point 
source with a power 
law (PL) spectrum, $dN/dE = N_0 (E/E_0)^{-\Gamma}$, where $E_0$ was fixed 
at the catalog value 1018.59 MeV, and the \gr\ counterpart to NGC~253
was modeled as a point source with a log-parabola (LP) spectral form,
$dN/dE = N_0 (E/E_b)^{-[\alpha +\beta\ln(E/E_b)]}$, where $E_b$ was fixed at
the catalog value 1048.36\,MeV.
\begin{figure*}
	\centering
\includegraphics[width=0.32\textwidth]{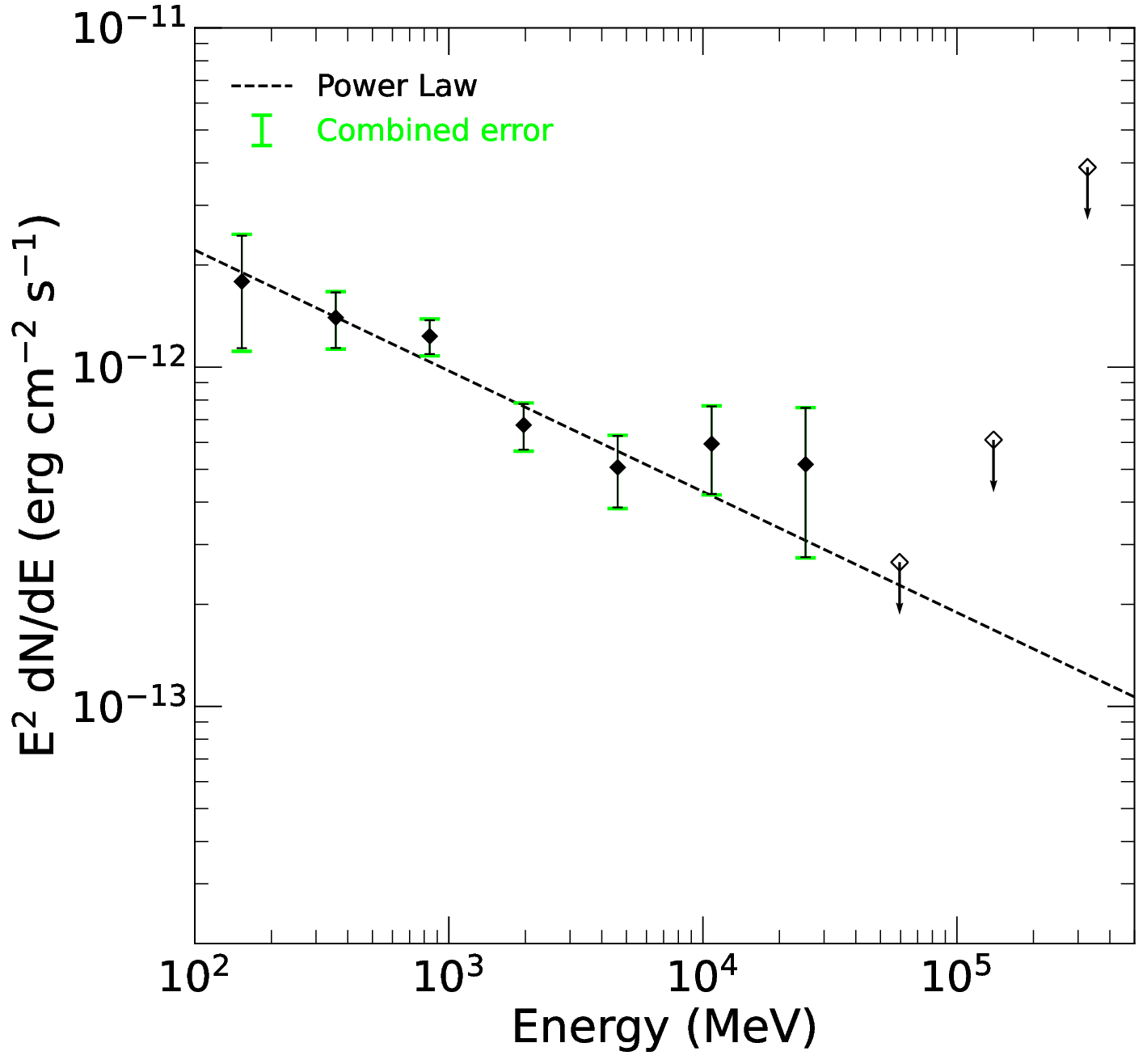}
\includegraphics[width=0.32\textwidth]{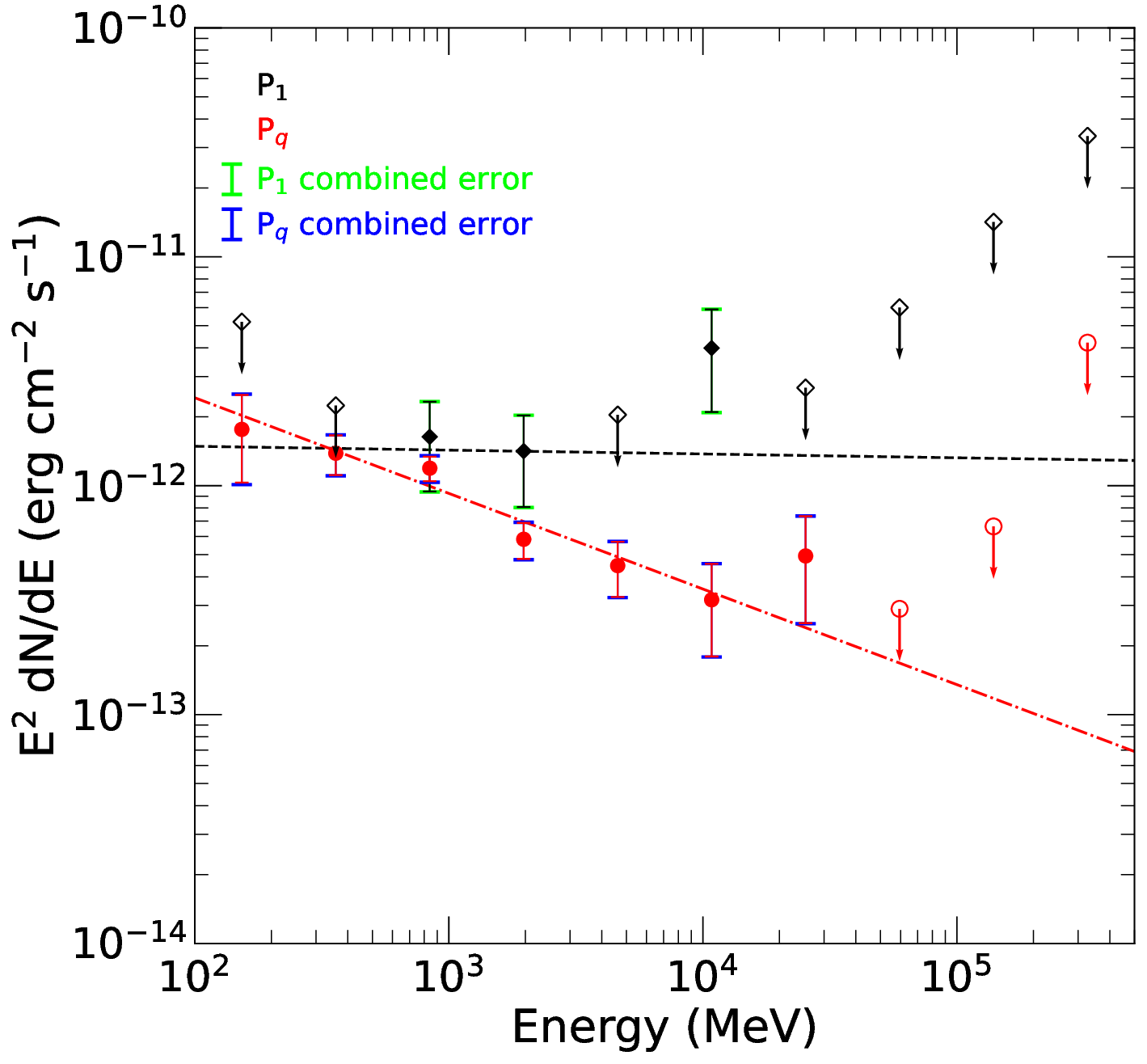}
\includegraphics[width=0.32\textwidth]{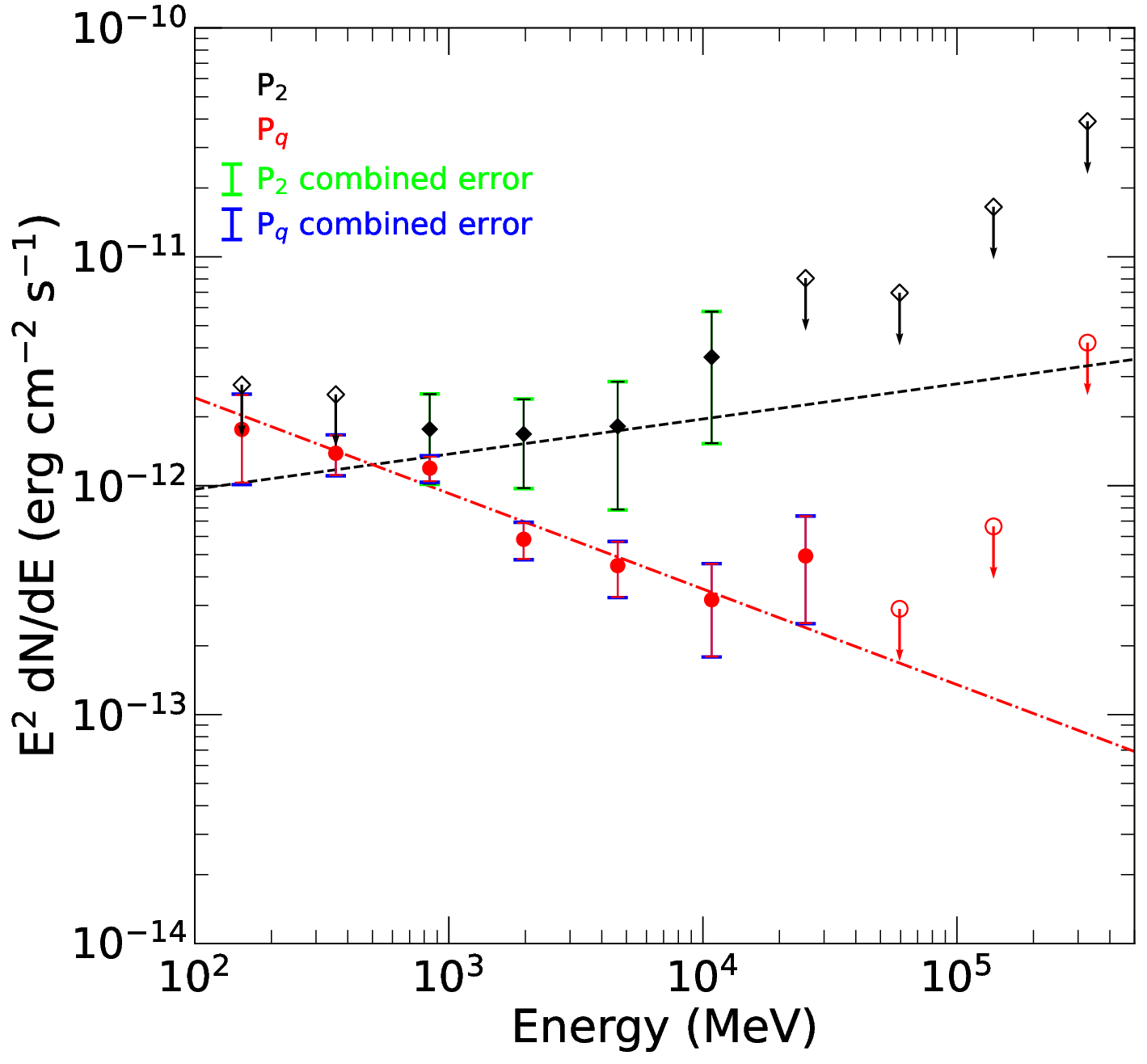}
	\caption{{\it Left:} \gr\ spectrum of \ngc\ in 0.1--500\,GeV. The PL 
	fit (dashed line) is shown for comparison. {\it Middle} and 
	{\it Right:} \gr\ spectra (black) of \ngc\ in P$_1$ and P$_2$, as 
	compared to the spectrum in P$_q$ (red). The PL fits to the data in
	the respective time periods are also shown (marked red for P$_q$ and
	black for P$_1$ and P$_2$).}
\label{fig:spec}
\end{figure*}

\subsection{Analyses for NGC~1068}

\subsubsection{Likelihood Analysis and Spectrum Extraction}
\label{sssec:ls}

Setting a PL for \ngc, we performed the standard binned likelihood analysis 
to the data in 0.1--500\,GeV. The obtained results, given in Table~\ref{tab:ts},
are consistent with the catalog values.  
%%Because in the following sections we analyzed the 2--500\,GeV data, we also performed the likelihood analysis to the data in the energy range.  The obtained results are also given in Table~\ref{tab:ts}.

We then extracted the $\gamma$-ray spectrum of NGC~1068 by performing maximum 
likelihood analysis to the LAT data in 10 evenly divided energy bins in 
logarithm from 0.1 to 500\,GeV. In the extraction, only the spectral 
normalizations of the sources within 5\arcdeg\ from NGC~1068 were set free, 
and all other spectral parameters of the sources in the source model were 
fixed at the best-fit values obtained from the likelihood analysis.
For the obtained spectral data points, we kept those with
TS$\geq$4 and derived the 95\% flux upper limits otherwise. The spectrum and
the PL model fit are shown in Figure~\ref{fig:spec}. For the spectrum,
the systematic uncertainties given in Section~3.5 of \citet{4fgl},
which are 10\% in 100--300\,MeV and 5\% in the higher energy ranges,
were added in quadrature to the statistical ones. 
We noted that while the
spectrum can be relatively well described with the PL model, 
the spectral data points with energies approximately greater than 2\,GeV
appear flat, which provided a hint for our following analysis.
\begin{figure}
\centering
	\includegraphics[width=0.42\textwidth]{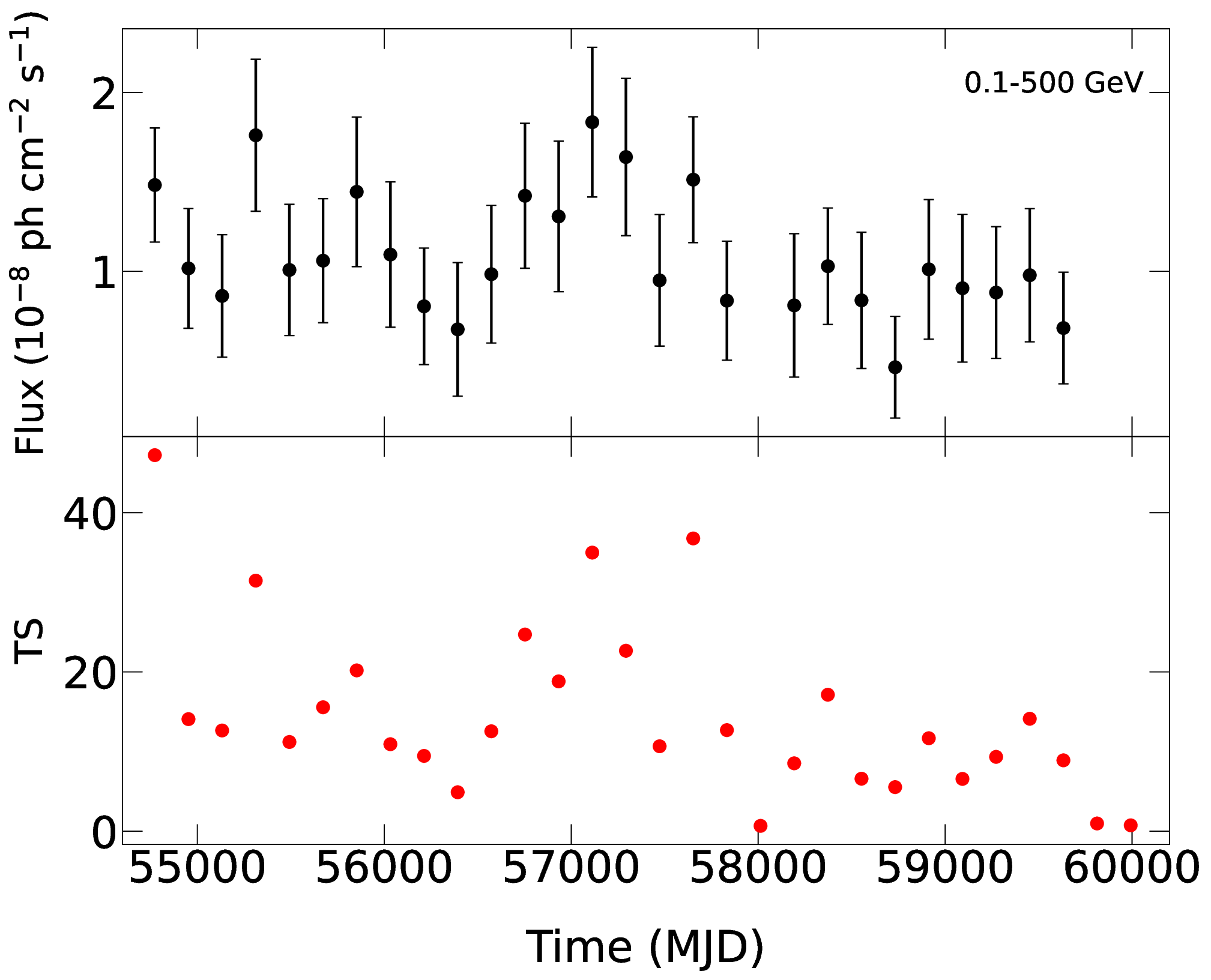}
	\includegraphics[width=0.42\textwidth]{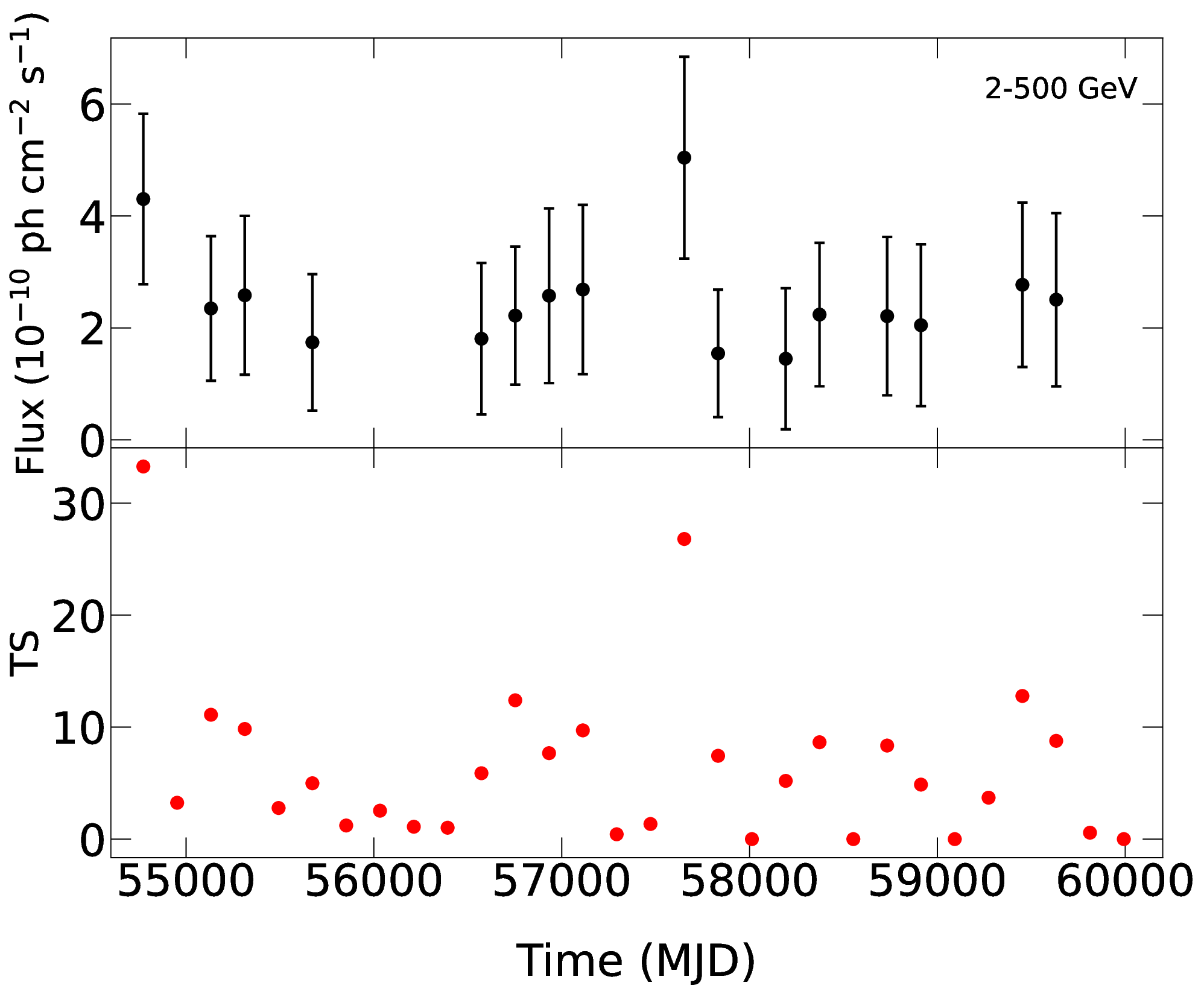}
	\caption{180-day binned light curves and TS curves
	of NGC~1068 in 0.1--500 GeV ({\it top} panels) and in 
	2--500\,GeV ({\it bottom} panels). Fluxes with TS$\geq$4 are kept 
	in the light curves.
	\label{fig:lc}}
\end{figure}

\subsubsection{Light curve analysis} 
\label{sssec:lc}

We extracted the source's 0.1--500\,GeV light curve by setting 180-day a time 
bin and performing the maximum likelihood analysis to each set of the
time-bin data. In the extraction, only the normalization parameters of 
the sources within 5\arcdeg\ from NGC~1068 were set free and
the other parameters were fixed at the best-fit values obtained from 
the above maximum likelihood analysis. The extracted light curve, for
which a systematic uncertainty of 2\% (see, e.g., \citealt{aje+20}) was added,
is shown
in the top panels of Figure~\ref{fig:lc}. The light curve shows possible
variations, but when we used the variability index TS$_{var}$ \citep{nol+12}
to evaluate, no significant variations were found as TS$_{var}\simeq 37.5$,
lower than the threshold value 49.6 (at a 99\% confidence level, for 29
degrees of freedom).

Given the spectrum of the source and our suspicion about the spectral
flatness above 2\,GeV, we also extracted a 2--500 GeV light curve 
(bottom panels of Figure~\ref{fig:lc}) with the same setup as the above 
and the same systematic uncertainty added. 
In this high energy range, there are two data points (especially the 
corresponding TS ones), at the beginning $\sim$MJD~54770
and the middle $\sim$MJD~57650, possibly indicating two flaring events.
However by calculating TS$_{var}$, we noted that the light curve
did not show significant variations, as TS$_{var}\simeq 29.5$.

\begin{figure}
\centering
	\includegraphics[width=0.46\textwidth]{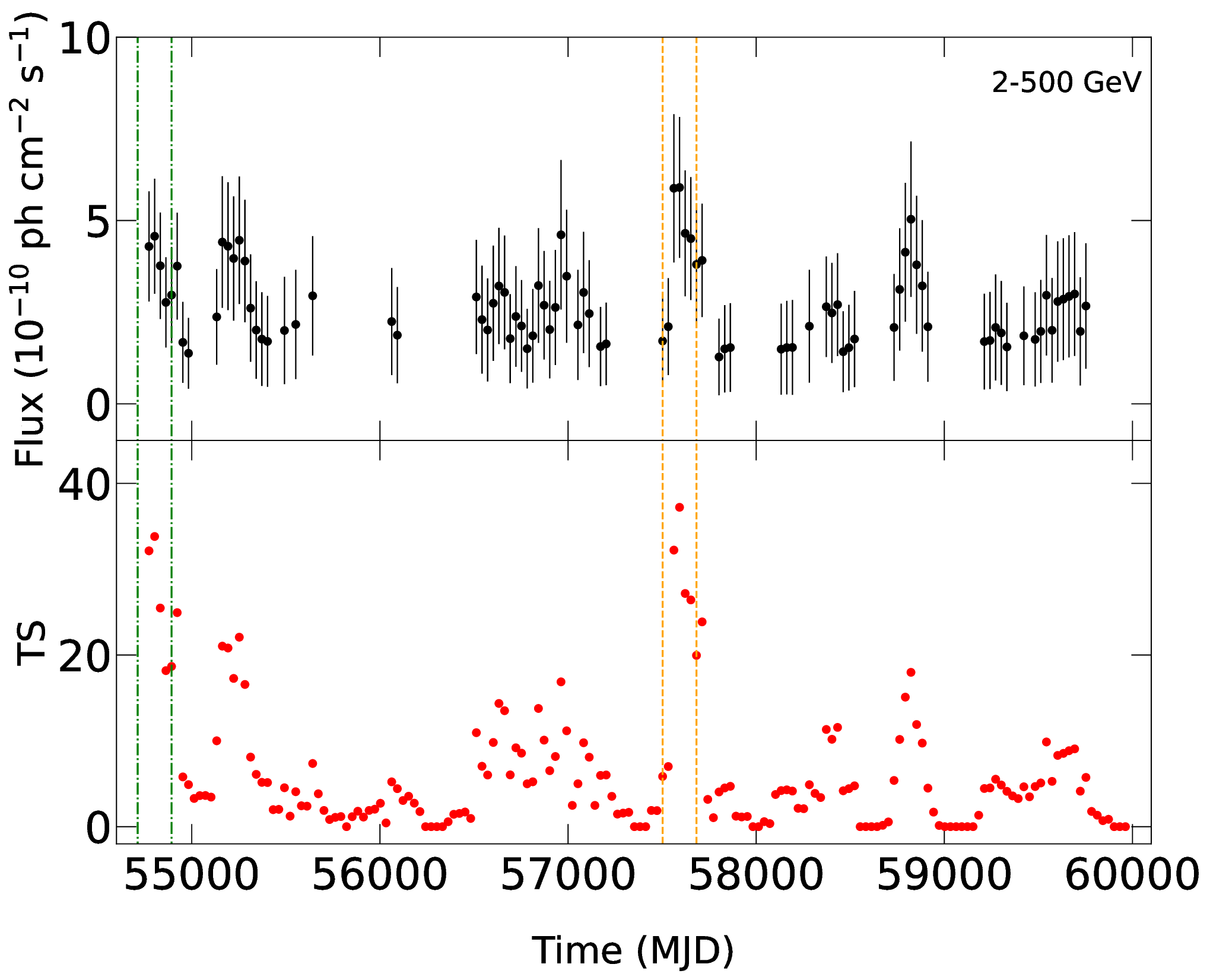}
	\caption{180-day binned smooth light curve ({\it top}) and TS curve
	({\it middle}). The time periods P$_1$ and P$_2$, during which the TS
	values reach to $\sim$40, are marked by green dash-dotted and 
	yellow dashed
	lines respectively. }
	\label{fig:slc}
\end{figure}

In any case, we refined the time periods of the two possible events 
by obtaining a smooth light curve, for which a time bin of 180\,day but with
a moving step of 30\,day was used. The obtained smooth light curve (with
a 2\% systematic uncertainty added) and TS 
curve are shown in Figure \ref{fig:slc}.  The two time bins with the highest TS 
values were thus found to be MJD~54712.7--54892.7 and MJD~57502.7--57682.7. 
We defined the first and second time bins as P$_1$ and P$_2$, respectively,
and the rest of the time periods as the `quiescent' P$_q$. 

\subsubsection{Likelihood and Spectral Analysis for P$_q$, P$_1$, and P$_2$}
\label{subsec:per}

We performed likelihood analysis to the data in P$_q$, P$_1$, and P$_2$,
while a PL spectral model was still used for \ngc. In this analysis,
the spectral parameters of the sources within 5\arcdeg\ from the target
were set free and all the other parameters (except the normalizations of
the two background components) were fixed at their catalog values.
The obtained results are given in Table~\ref{tab:ts}. As can be seen,
the emissions in P$_1$ and P$_2$ ($\Gamma\simeq 2.02$ and 1.85 respectively)
were harder that that in P$_q$ ($\Gamma\simeq 2.43$). The significances for
the spectral differences were $\simeq$2.2$\sigma$ and 3.2$\sigma$ respectively.
\begin{figure}
	\centering
\includegraphics[width=0.45\textwidth]{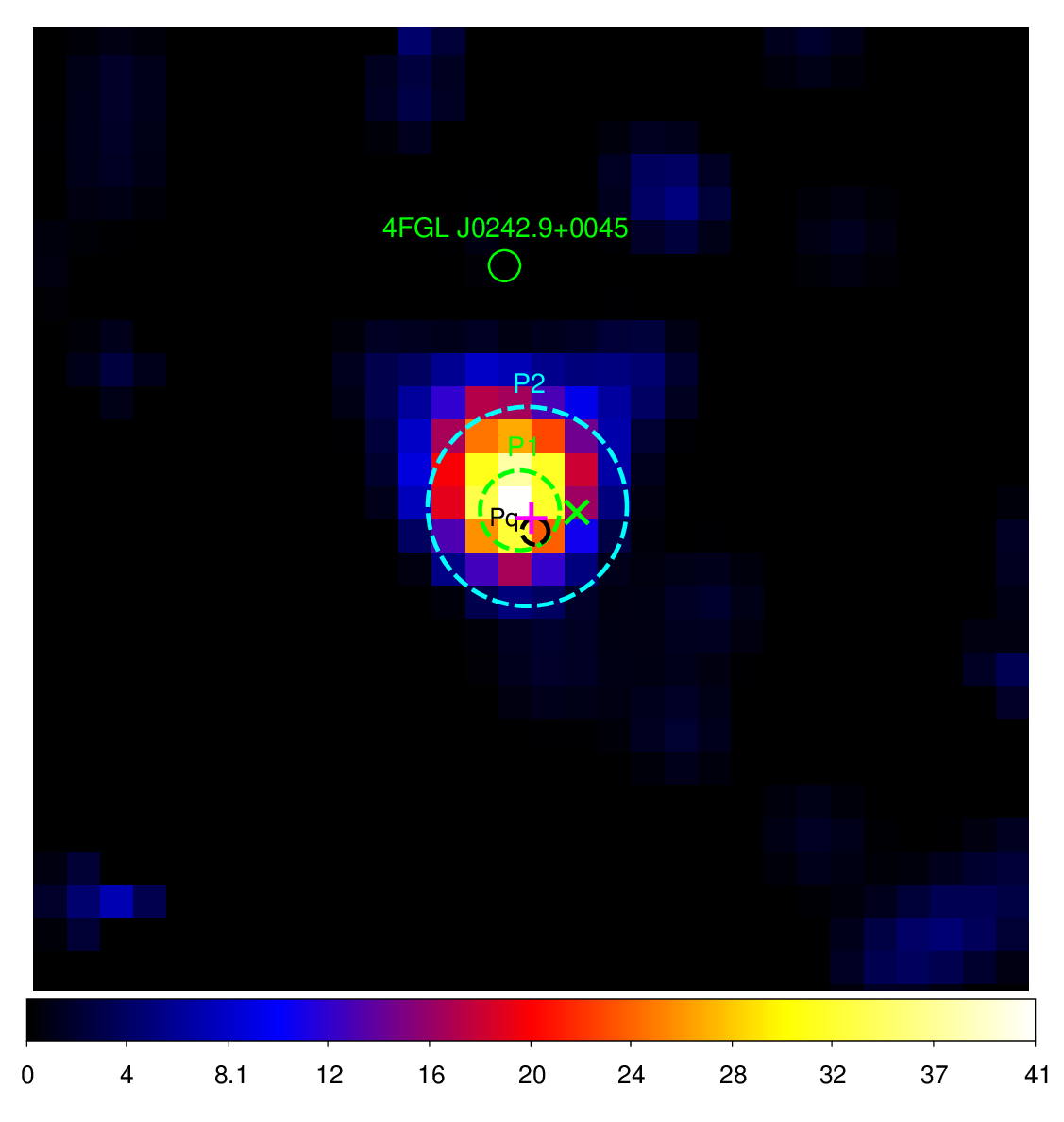}
	\caption{TS map of a 3\arcdeg$\times$3\arcdeg\ 
	region centered at NGC~1068
	in $\geq2$\,GeV energy range during the P$_1$ time period. 
	The positional error circles
	(2$\sigma$ values, which are 0\fdg12, 0\fdg3, and 0\fdg04)
	determined in P$_1$, P$_2$, and P$_q$ time periods,
	respectively, are marked as dashed green, cyan, and black circles.
	The pink plus marks the position of NGC~1068 and the green cross
	marks the position of [vv2000] J024207.2+000038, a nearby quasar 
	example.
	\label{fig:1068}}
\end{figure}

We extracted spectra from the data during the three time periods to check
the likelihood results. The same setup and procedures as that in 
Section~\ref{sssec:ls} were used.  We kept the spectral data points 
with TS $\geq$ 4 and showed the derived 95$\%$ flux upper limits otherwise. 
The same systematic uncertainties mentioned in Section~\ref{sssec:ls}
were added.
The obtained spectra are shown in Figure~\ref{fig:spec}, with the spectra of
P${_1}$ and P${_2}$ compared to that of P${_q}$ respectively. As can be seen,
the spectral models from the likelihood analysis
can generally describe the spectra, while only a $\sim$10\,GeV spectral data
point appears slightly deviating away from the model fit in P$_1$.

For the P$_1$ and P$_2$ two possible spectral change events, 
we carefully
checked for other possibilities, such as if they could be induced by
the proximity to the Sun, the solar flares\footnote{http://hesperia.gsfc.nasa.gov/fermi\_solar/} or gamma-ray bursts \citep{grbs} 
in the source field, or flares of background quasars during the time periods.
The first two possibilities were excluded as no such cases occurred in 
the RoI during
the time periods, which was confirmed by non-detection of short flaring events
in such as 1-day--binned light curves we tested to construct. 
For the third one, there are many quasars,
given in the SIMBAD database, that are located $\sim$2\arcmin--30\arcmin\ to 
NGC~1068,
for example [VV2000] J024207.2+000038 (\citealt{vv00}; see 
Figure~\ref{fig:1068}).  
However, there are no reports of this source being a blazar.

In addition, there are 9 sources listed in 4FGL-DR3 within 5\arcdeg\
of NGC~1068, among which the closest one 4FGL~J0242.9+0045 
(Figure~\ref{fig:1068}) is 0\fdg77 away. 
Among them, six were within 4\arcdeg\ and fainter than our target. We checked 
if any of
them could be variable during P$_1$ and P$_2$ by performing binned likelihood
analysis to the data for each of them. No significant variations were seen,
as the spectral parameters of them during either P$_1$ or P$_2$ were consistent
within uncertainties with those obtained from the likelihood analysis to 
the whole data. We also determined the positions of NGC~1068 during P$_1$,
P$_2$, and P$_q$ in $\geq 2$\,GeV energy by running {\tt gtfindsrc} 
in Fermitools, and they are shown in Figure~\ref{fig:1068}, a TS map
calculated from the $\geq 2$\,GeV data during the P$_1$ time period. 
As can seen, the positions are
compatible, excluding the possibility that the apparent hardening of NGC~1068 
was instead caused by spectral changes or flares of some nearby sources.

\subsubsection{Summary of the results}

By checking the $\geq 2$\,GeV emission of \ngc, we found two half-year
periods, during which the TS values appeared larger than those of the rest. 
The corresponding spectra were found to become harder from 2\,GeV than 
the spectrum in quiescence, i.e., having more higher energy photons, and in one
time period, the deviation of the spectrum from the latter reached 3.2$\sigma$.
However, there were trials to be considered in our analysis, 
such as the used time bin and energy bin in light curves and spectra 
respectively. As we did not particularly test
to find the optimal values for them, where they were approximately chosen 
based on the TS value of the source, it is hard to evaluate a trial number for 
our analysis. Also an energy range of $\geq$2\,GeV was chosen
based on the visual inspection of the source spectrum.
If we take a trial number of
$\geq$100 considering all the numbers of time bins and energy bins, the 
global significance is largely reduced to be $\sim$0.
Thus we conclude that only possible and interesting
spectral changes in the $\geq 2$\,GeV energy range were found for \ngc.
\begin{table}
	\centering
        \caption{Likelihood analysis results for NGC~253}
        \label{tab:like}
        \begin{tabular}{lcccc}
        \hline
        \hline
	Data/Model & $\alpha$/$\Gamma$ & $\beta$ & $-\log(L)$ & TS \\
          \\
        \hline
      Catalog/LP   &  2.03$\pm$0.07 & $0.10\pm 0.04$ &  & 736 \\
		Whole/LP   & $2.07\pm 0.07$ & $0.07\pm 0.03$ &  & 896\\\hline
%%		$\geq5$ GeV/PL &   $2.26\pm0.15$  &     &   &  166  \\\hline
		P1/LP & 2.08$\pm$0.08 & 0.03$\pm$0.04 & 441980.4 & 436 \\
		P1/PL & 2.11$\pm$0.06 &    & 441980.8 & 437\\\hline
		P2/LP & 1.94$\pm$0.16 & 0.17$\pm$0.08 & 481222.6 & 454\\
		P2/PL & 2.22$\pm$0.06 &    & 481226.3 & 480\\\hline
\end{tabular}
\end{table}

\subsection{Analyses for NGC 253}
\subsubsection{Likelihood and spectral analysis}
\label{sec:ls3}

Using the source model described in Section~\ref{ssec:data}, we performed 
the standard binned likelihood analysis to the data in 0.1--500\,GeV.
The obtained results, given in Table~\ref{tab:like}, are consistent
with the values given in 4FGL-DR3, while it can be noted that $\beta$ is quite
small, suggesting a PL model could likely provide an equally good fit.
%Because in the following sections we analyzed the 5--500\,GeV data, we also performed the likelihood analysis to the data in this energy range, where a PL was assumed. The results are also provided in Table~\ref{tab:like}.

\begin{figure}
	\centering
\includegraphics[width=0.42\textwidth]{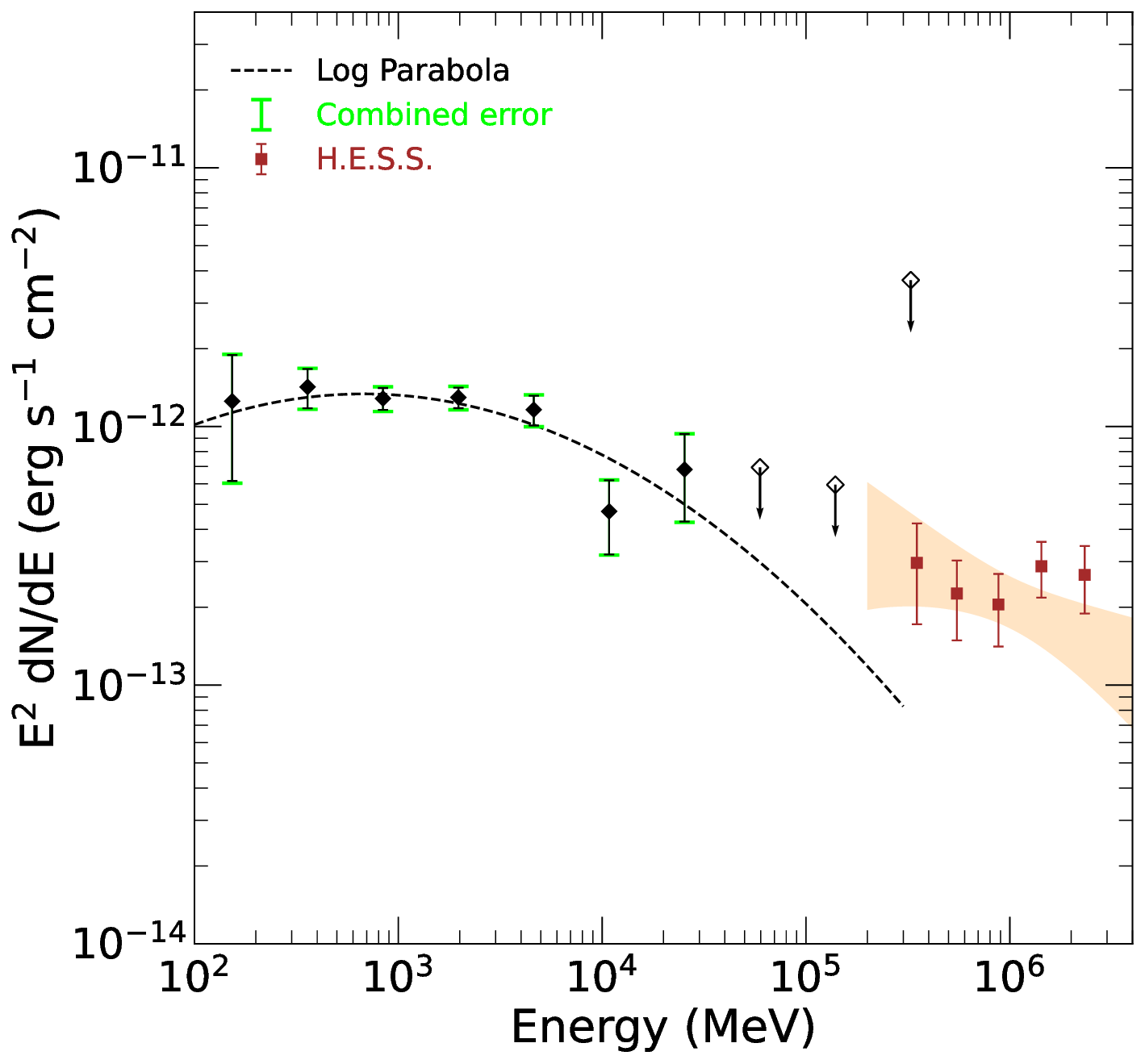}
\includegraphics[width=0.42\textwidth]{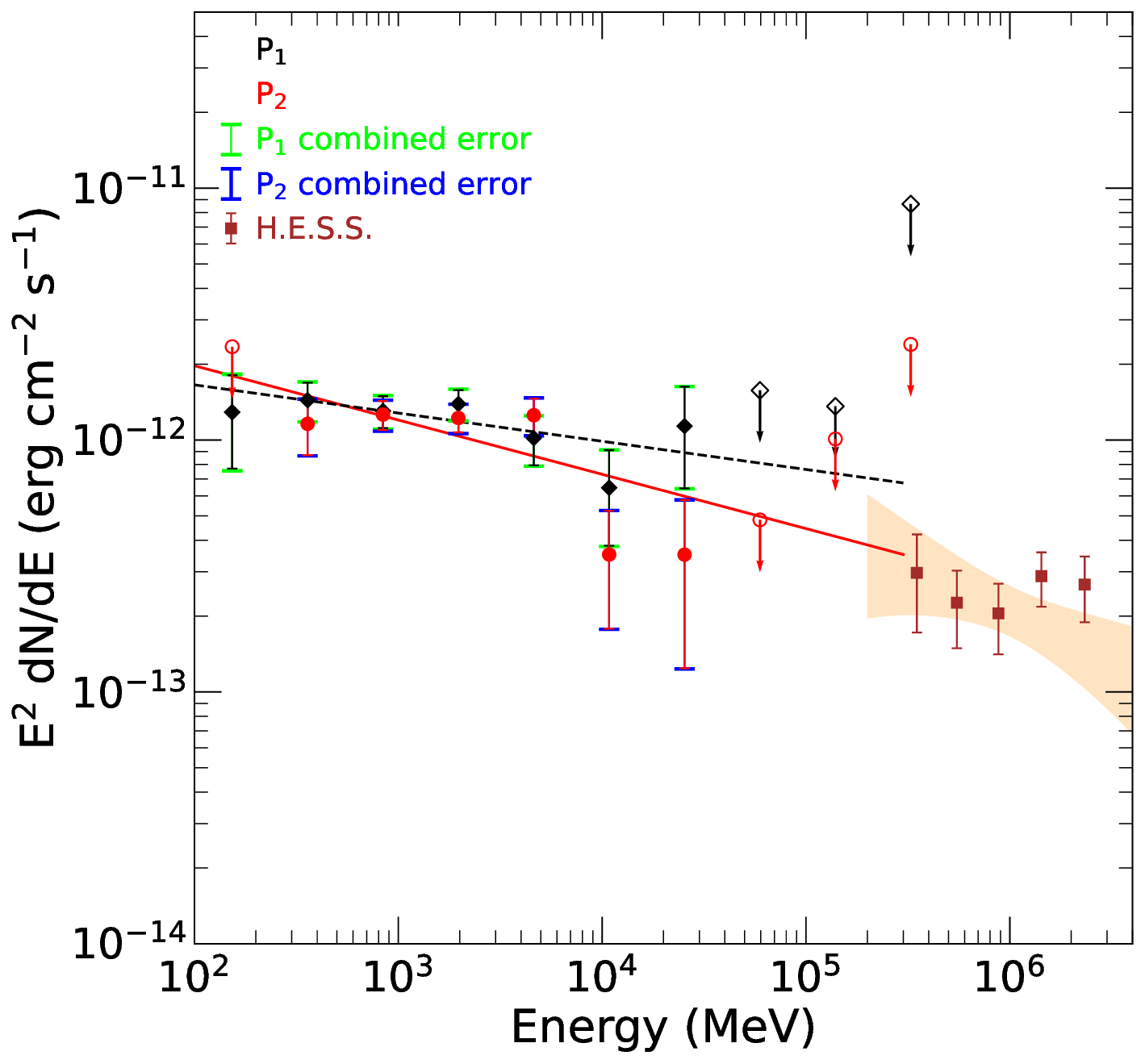}
	\caption{{\it Top:} \gr\ spectrum of NGC 253 in 0.1--500\,GeV. The LP
	fit (dashed line) is shown for comparison. {\it Bottom:} 
	\gr\ spectra of NGC~253 in P1 and P2 (black and red respectively), 
	with the corresponding best-fit spectral models also shown. The HESS
	VHE spectral data points (brown) and spectral shape (yellow shaded
	region; \citealt{hess18}) are also shown.
	}
\label{fig:spec3}
\end{figure}

We then extracted the $\gamma$-ray spectrum of NGC~253 by performing
maximum likelihood analysis to the LAT data in 10 evenly divided energy bins
in logarithm from 0.1 to 500\,GeV. 
The same setup and procedure as those in Section~\ref{sssec:ls} were used.
For the obtained spectral data points, we kept those with
TS$\geq$4 and uncertainty/flux $\leq$ 50\%, and derived the 95\% flux upper
limits otherwise. The spectrum (with the systematic uncertainties 
added) and
the LP model fit are shown in Figure~\ref{fig:spec3}. We noted that while the
spectrum can be relatively well described with the LP model, there is a
possible flux drop-off at $\gtrsim$5\,GeV, which provided a hint for our
following analyses.

\subsubsection{Light curve analysis}
\label{subsec:lc3}

\begin{figure}
\centering
	\includegraphics[width=0.42\textwidth]{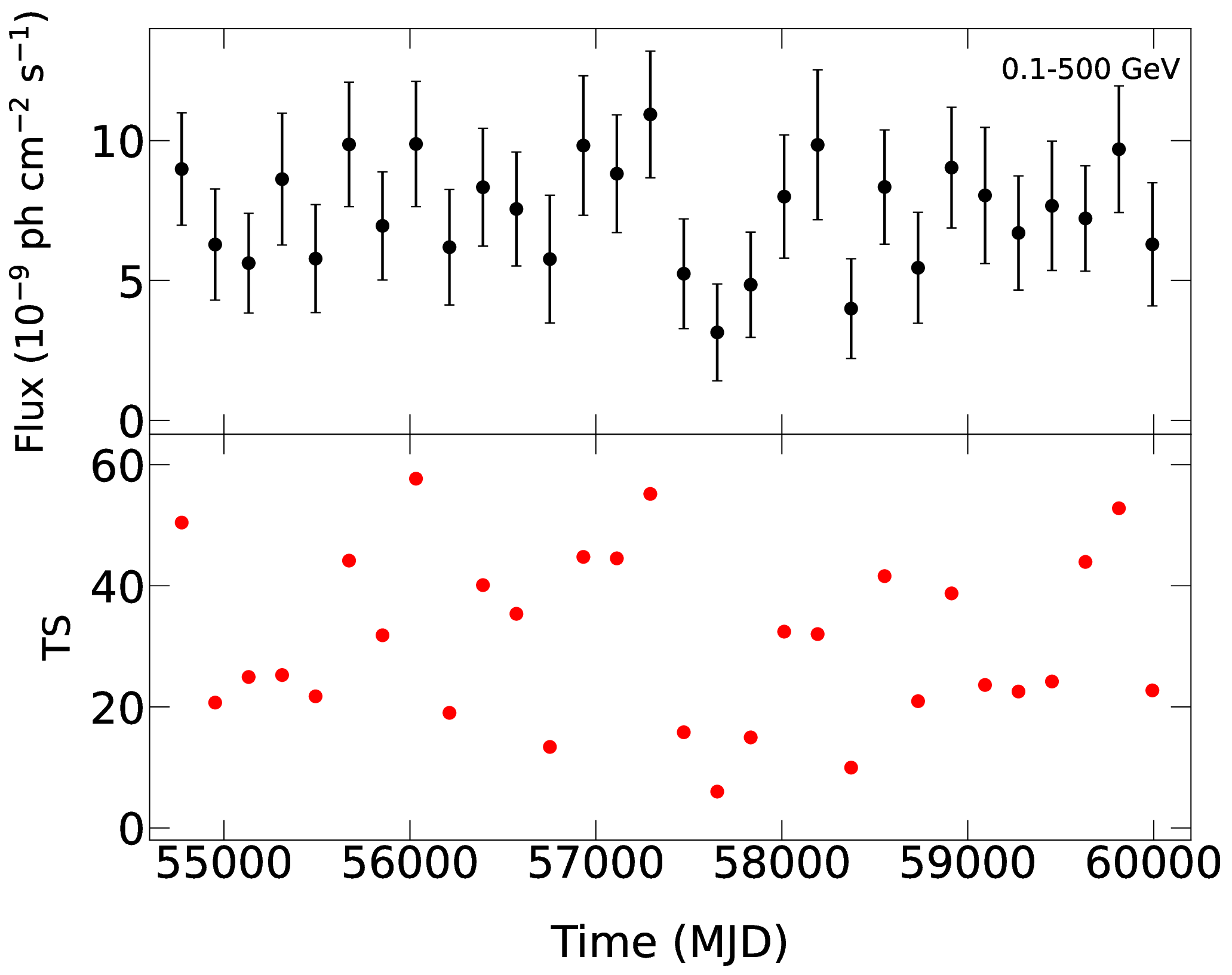}
	\includegraphics[width=0.42\textwidth]{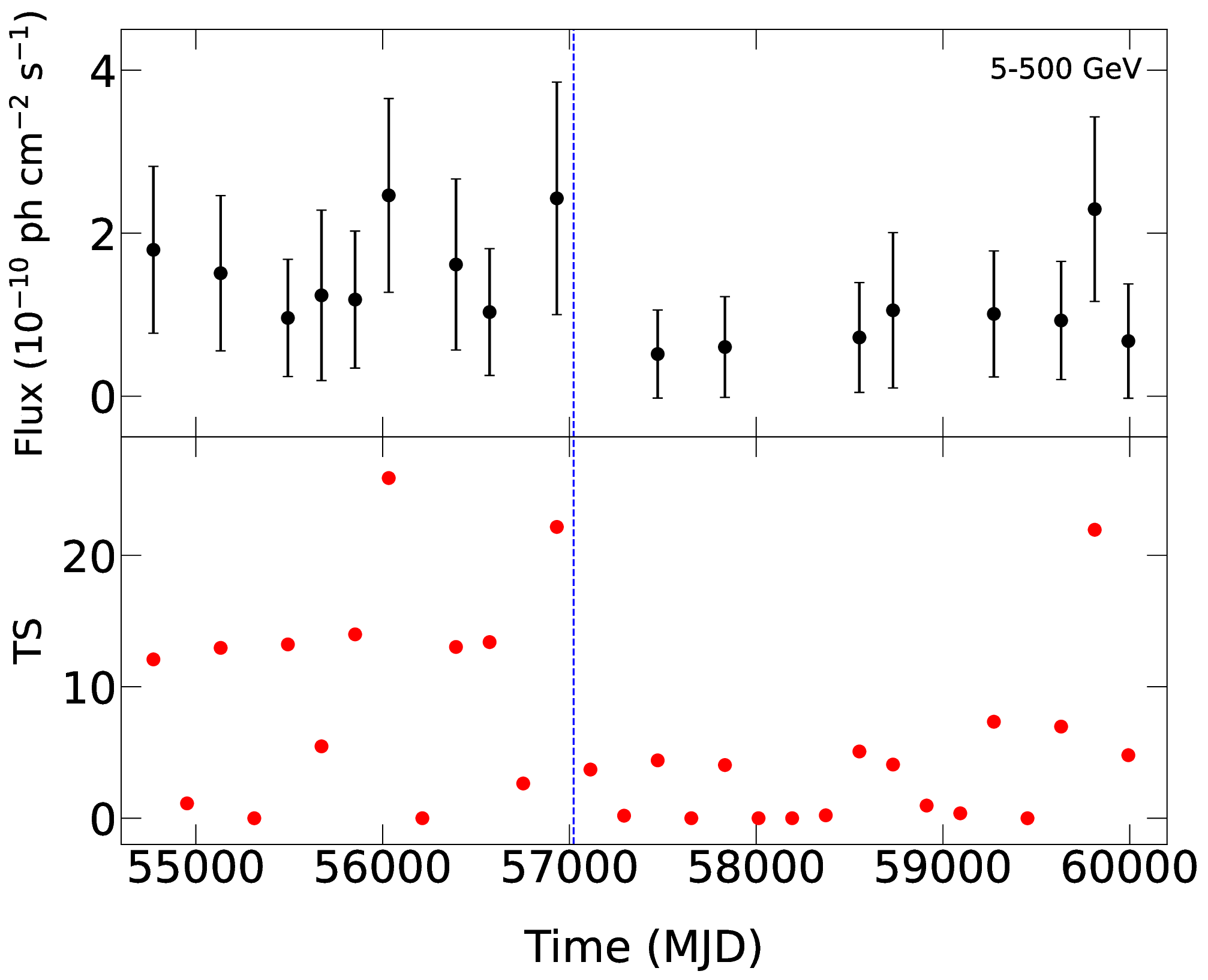}
	\caption{180-day binned light curves and TS curves of NGC~253 in
	0.1--500 GeV ({\it top} panels) and 5--500\,GeV ({\it bottom} panels).
	Fluxes with TS$\geq$4 are kept in the light curves. In the bottom
	panels, a blue dotted line indicates MJD~57022.7, an approximate time
	for the flux changes. 
	\label{fig:lc3}}
\end{figure}

We extracted the source's 0.1--500\,GeV light curve using the same setup
and procedure described in Section~\ref{sssec:lc}.
The extracted light curve is shown in Figure~\ref{fig:lc3}, for which
the 2\% systematic uncertainty was added.
The variability index TS$_{var}$ \citep{nol+12} was determined to be
TS$_{var}\simeq 24.9$, lower than the threshold value 49.6
(at a 99\% confidence level for 29 degrees of freedom). Thus no variability
was found in the whole data.

Given the spectrum of the source and its possible spectral
drop-off above 5\,GeV, we also extracted a 5--500 GeV light curve
(bottom panels of Figure~\ref{fig:lc3}) with the same setup as the above
and the same systematic uncertainty added.
In this high energy range, TS values of the data points
seem to have decreased after $\sim$MJD~57000.
However TS$_{var}\simeq 30.7$ was obtained for the light curve,
not indicating significant variations.

In any case, we defined the two time periods based on the light curve,
phase 1 (P1) in MJD~54682.7--57022.7 and phase 2 (P2) in
MJD~57022.7--60040.0.

\subsubsection{Likelihood and spectral analysis for P1 and P2}
\label{subsec:pha}

In order to check possible property differences between the two time periods,
we first performed likelihood analysis to their data.
For the target, we tested both LP and PL spectral models.
The results are summarized in Table~\ref{tab:like}. We found that
in P1 and P2, the PL was preferred given the same likelihood values
obtained from the both models.
The obtained $\Gamma$ values are consistent with each others within 
uncertainties.

Spectra of P1 and P2 were extracted from the data during the two time periods.
The setup and procedure were the same as in Section~\ref{sssec:ls}, and
the systematic uncertainties were included in the results.
With the spectra shown in Figure~\ref{fig:spec3}, it can be seen that
no significant flux changes could be determined due to the large
uncertainties.

\begin{figure*}[!ht]
\centering
\includegraphics[width=0.32\textwidth]{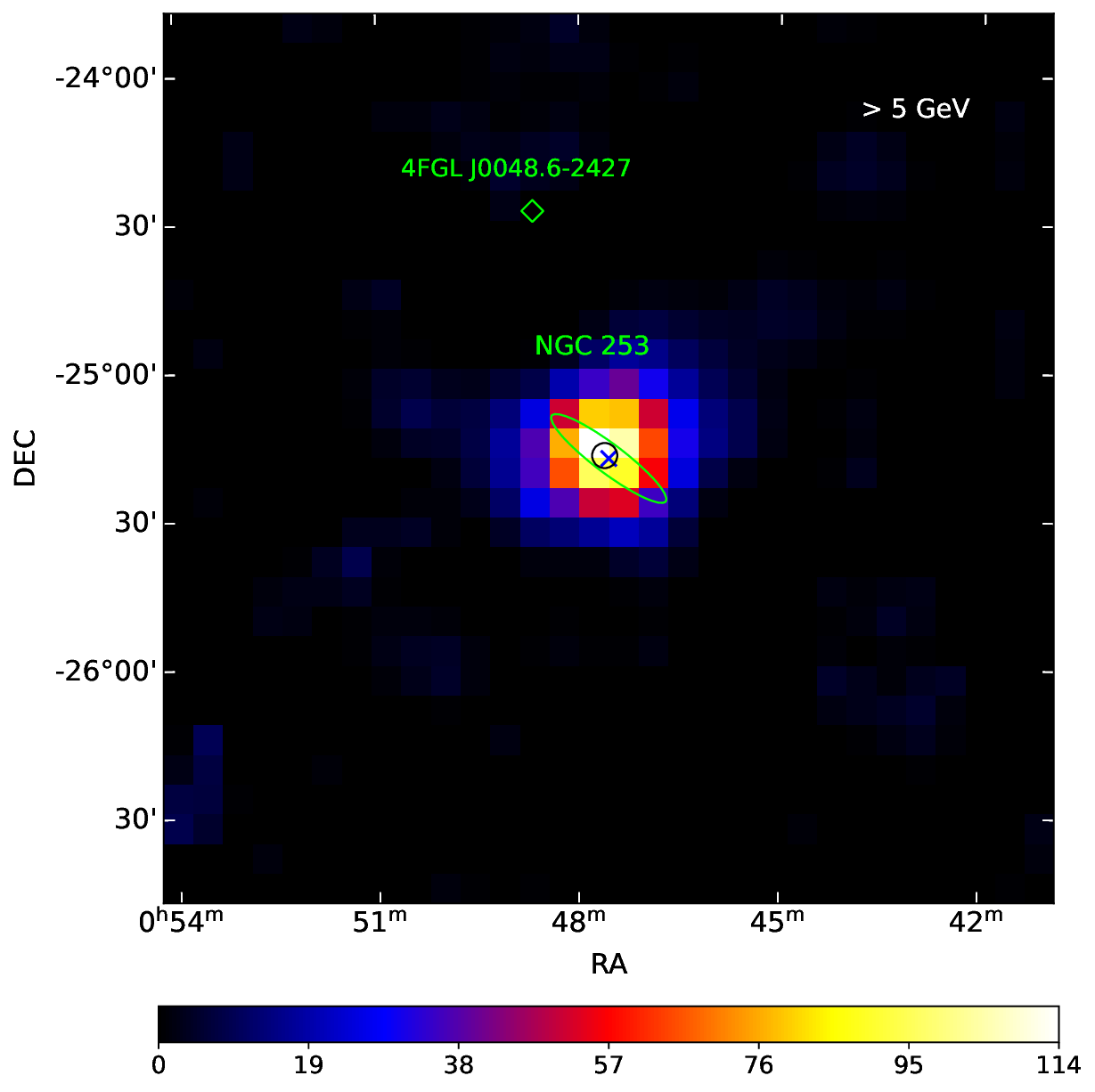}
\includegraphics[width=0.32\textwidth]{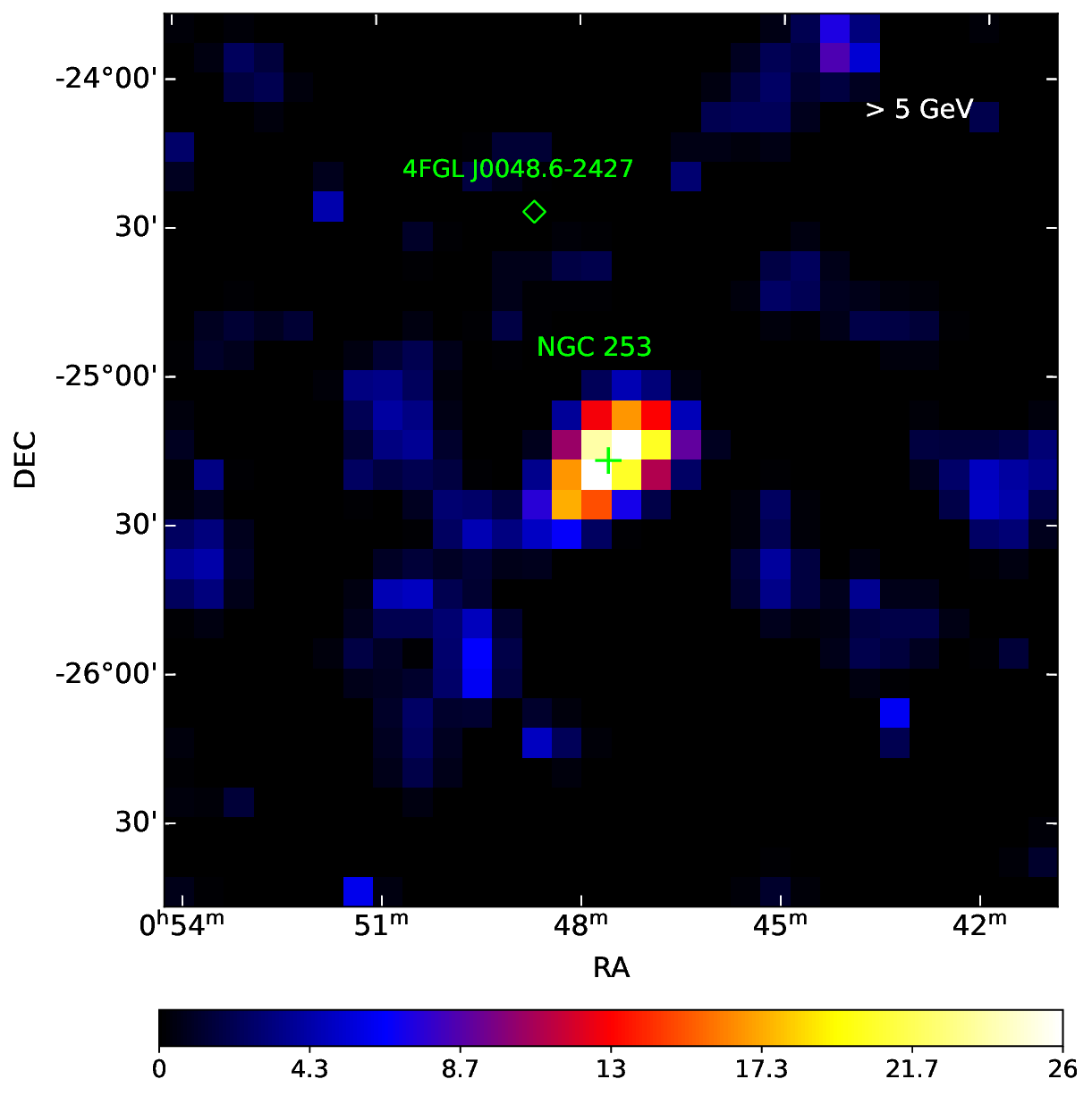}
\includegraphics[width=0.32\textwidth]{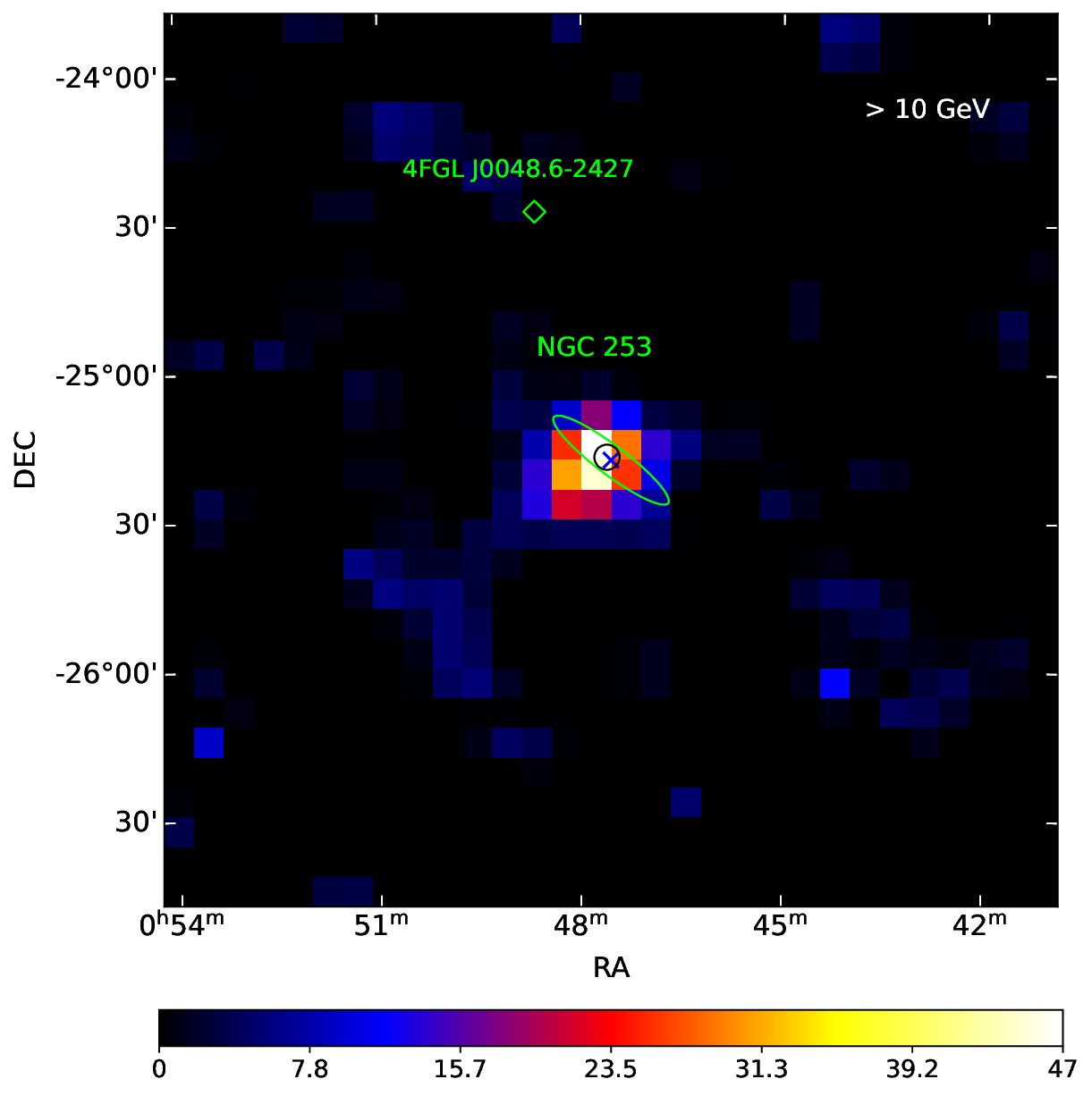}
\caption{TS maps of a $3\arcdeg\times 3\arcdeg$ region centered at NGC~253
	in energy ranges of 5--500\,GeV during P1 ({\it left}) and
	P2e ({\it middle}) and of 10--500\,GeV during the whole time period
	({\it right}). The optical shape of the galaxy is shown as a green
	ellipse, with the center marked as a cross. The circle in the left
	and right panels marks the 2$\sigma$ error circle determined for
	the 5--500\,GeV emission of the galaxy.
	\label{fig:tsmap}}
\end{figure*}

We thus calculated the TS maps centered at NGC~253 in energy range of $>$5\,GeV
during P1 and MJD~57022.7--59362.7 (hereafter P2e), where the second one
was set such that its length is equal to that of P1 for direct comparison. 
The TS maps
are shown in Figure~\ref{fig:tsmap}. As can be seen, the TS value is $\sim$114
in P1 while $\sim$26 in P2e.

We performed the likelihood analysis to the data in 5--500\,GeV in P1 and P2
(P2e), and obtained the fluxes of 11.8$\pm2.4 \times 10^{-11}$ and
5.6$\pm1.5 \times 10^{-11}$\,photon\,cm$^{-2}$\,s$^{-1}$ 
(4.1$\pm1.5\times 10^{-11}$\,photon\,cm$^{-2}$\,s$^{-1}$). The flux values
are consistent with the TS values as the flux in P2 or P2e is approximately 
half of that in P1. However, 
the flux difference between P1 and P2 is not significant
due to large uncertainties (similarly indicated by 
the spectral data points in Figure~\ref{fig:spec3}).

\subsubsection{Additional analyses}
\label{sssec:aa}

For discussing the \gr\ emission of NGC 253, we conducted additional analyses
and provide the results in this section.

First, we checked if the \gr\ emission could be extended  
or not. In this analysis, the whole 0.1--500\,GeV data were used. We set
a uniform disk model for the emission in the source model, with its radius
ranging from 0\fdg1 to 1\fdg0 at a step of 0\fdg1. Performing the likelihood
analysis, the likelihood values from different radii were obtained and 
compared to that from a point
source. The analysis and comparison indicated the emission was consistent 
with being from a point source.

Second, we determined the position of $\geq$5\,GeV emission of NGC 253, to
check if the high energy emission was consistent with being from the center
of NGC 253. Running {\tt gtfindsrc} to the data, we obtained 
R.A.=11\fdg90, Decl.=$-$25\fdg28 (equinox J2000.0), with a 2$\sigma$ 
uncertainty of 0\fdg04. The error circle contains the center's position of 
NGC 253 (Figure~\ref{fig:tsmap}). Note that given the LAT's $\sim$0\fdg25 
containment angle at 5\,GeV\footnote{https://www.slac.stanford.edu/exp/glast/groups/canda/\\lat\_Performance.htm}, 
smaller than the major axis of the galaxy, the emission should
arise from the central region. To further check this, we calculated a 
$\geq 10$\,GeV TS map (Figure~\ref{fig:tsmap}), at which energy the containment
angle is $\lesssim$0\fdg2. The TS map shows that the emission appears
smaller than the major axis of the galaxy.

\subsubsection{Summary of the results}

By checking the emission of NGC 253, we found that 
its $\geq$5\,GeV part had a possible flux drop around MJD~57023, 
with the flux before the date being $\sim$2 times that
after the date. This change is revealed by the TS maps calculated 
for the two time periods. However the change was not significant as
the flux uncertainties, particularly that of the second part, are large.
Thus we only found a detection significance decrease for NGC~253.
In addition, the \gr\ emission
in 0.1--500\,GeV is consistent with being a point source, and the high-spatial
resolution TS maps obtained at higher, $\geq$5\,GeV or $\geq$10\,GeV energies
indicate that the emission likely arises from the central region of NGC 253.

\section{Discussion}
\label{sec:dis}

For the purpose of fully studying the \gr\ emissions of the nearby galaxies 
NGC~1068 and NGC~253, which can potentially help improve our understanding of
the physical processes in them, we carried 
out detailed analysis.
In their \gr\ emissions, we found possible spectral or flux variations at
$\geq$2\,GeV or $\geq$5\,GeV high-energy ranges. Below we discuss 
possible implications of the findings respectively for the two galaxies.
%%Because of the presence of the variable components, additional processes are likely required in order to explain the \gr\ emissions from them. For \ngc, we suggest emission from its jet may contribute to the observed. It is not clear about the nature of $\geq$5\,GeV variations seen in NGC~253, but the variations may be related to its potential AGN.

\subsection{NGC 1068}

From the beginning when \gr\ emission was detected from \ngc, it has been 
realized that the emission is more intense than that estimated from typical 
starburst-induced \gr\ radiation models \citep{len+10,yoa+14,eb16}.
Thus alternative possibilities have been proposed.  \citet{len+10} used a jet 
model, as a kpc-scale radio jet is seen as a component of the 
galaxy \citep{wu83}, although recently reported multi-epoch
Very Long Baseline Array radio observations of NGC~1068 actually question
the presence of (small-scale) radio jets \citep{fis+23}.
\citet{lam+16} considered the AGN-driven outflow
\citep{kri+11} that induces shocks to produce high-energy particles. 
Indeed, AGN-driven outflows have been reported to be detected in $\gamma$-rays
\citep{aje+21}.

Recently to explain the observed neutrino emission, high-energy particles 
are suggested to be produced in
the corona around the central supermassive black hole \citep{mkm20,ino+20}. 
While these particles emit $\gamma$-rays in the same processes that produce 
the neutrinos, the optical depth to the high-energy photons is high 
(e.g., \citealt{mur22}) and thus the photons can not escape the central region 
until
further cascading to MeV energies. By including the processes of the corona
with those of the starburst, \citet{eic+22} provided a fit to both \gr\ and
neutrino spectra, while the \gr\ one was dominantly contributed by 
the starburst component. Alternatively, \citet{inos+22} proposed disk winds
as the power source for the neutrino and \gr\ emissions, with the latter 
arising from the interaction of the outgoing wind with the surrounding torus.
We note that \citet{amm23} very recently conducted analysis of
the \gr\ emission of \ngc, mainly by
extending the spectrum to as low as 20\,MeV. They argued that the low-energy,
$\leq$500\,MeV part of the emission comes from the corona and the other 
part from the starburst.

Our results also suggest that the \gr\ emission probably contains multiple 
components. As the typical 
starburst models generally work, as proved from studies of other starburst 
galaxies (e.g., \citealt{len+10,yoa+14,eb16}), a component from the starburst
should be present. Then at $\gtrsim$2\,GeV energies, our finding of 
spectral change events suggests there is another component. 
Given the variations, it is
possibly from the jet (it is not clear whether the outflow scenario would
produce variable emission), since jets' emissions are well known to be highly
variable. In the model of \citet{len+10}, the \gr\ emission is due to
the external inverse Compton process (EIC), from a blob of plasma that has
a bulk Doppler factor $\delta_{\rm b}$ and a radius $r_{\rm b}$. Taking 
half a year as the variability time scale $\Delta t$, we may estimate 
the size of the blob, 
$\sim c\Delta t\delta_{\rm b}\simeq 0.15\delta_{\rm b}$\,pc, where $c$ is
the speed of the light. This size value
is smaller than that ($\simeq$6.5\,pc) estimated from the 
spectral-energy-distribution (SED) fitting in \citet{len+10}, but we should 
note that there are large uncertainties on parameters in the modeling and
the size value (or the variability time scale) is in a reasonable range 
for this type of the emission regions. 

For the two variation events, we have checked multiwavelength data to find
any possible correlated variations, in particular at X-rays. 
\citet{oh+18} reported a hard X-ray
light curve in 14--195\,keV obtained with the Burst Alert Telescope (BAT)
onboard {\it the Neil Gehrels Swift Observatory (Swift)}, and the light curve
ends at $\sim$MJD~56500. The mission Monitor of All-sky X-ray Image (MAXI;
\citealt{mat+09})
also detects the galaxy and provides a 2--20\,keV light curve. However no
correlated variations were found from these two light curves. We note that
\citet{mar+16} reported a transient flux excess in $\geq 20$\,keV 
hard X-rays however in 2014 Aug. (before the second variation event),
and it was likely caused by a temporary decrease of the column density
of the obscuring material (in the torus) towards the AGN.
Future sensitive hard X-ray or MeV observations of NGC~1068 may help 
find hints for understanding the \gr\ properties, probing if there are 
related changes, as the emission in the bands
is suggested to be related to the \gr\ emission from the galaxy's central 
region \citep{mkm20,ino+20}.
%%We note that if the variable part in the \gr\ emission comes from the jet, radio monitoring of the jet might be able to provide supporting evidence by detecting similar variations.

\subsection{NGC 253}

As the starburst models can explain the observed GeV and VHE \gr\ emissions
from NGC~253 (e.g., \citealt{abd+10a,lac+11,abr+12,eb16,hess18}) and no flux 
variations
would arise from this sort of starburst-induced emission, it is unexpected
to find any variations probed in this work. Both the VHE detection
\citep{ace+09,abr+12} and our analysis results point the emission region at
the center of the galaxy. We thus suspect any possible flux variations,
as hinted by the TS value difference, are possibly related
to the central region, where its potential AGN has been 
searched (e.g., \citealt{wea+02,mul+10}).
From the work reported by \citet{leh+13}, we can see there are four relatively
bright X-ray sources in the center, with one of them (N-2003) suggested as
an AGN candidate. We checked the hard X-ray light curve of NGC~253 
reported from the {\it Swift}/BAT Hard X-ray Transient Monitor 
program \citep{kri+13}, but no significant variations could be
seen in the light curve. We also checked the X-ray data from targeted 
observations with the {\it Swift} X-ray Telescope (XRT), and no 
variation patterns
were seen in the flux measurements of the central X-ray source (which 
presumably consists of the four sources) before and 
after $\sim$MJD~57023 (2015 Jan. 1). There are also many sets of archival data 
from {\it Chandra} X-ray observations of NGC~253, but by checking
the data for the resolved X-ray sources, no clear related variations were
found.

Finally, we note that the VHE emission may be closely connected with the 
$\geq$5\,GeV part. In Figure~\ref{fig:spec3}, we show the HESS VHE data points
of the galaxy \citep{hess18}, which were obtained from the data collected
in years 2005 and 2007--2009.
The $\geq$5\,GeV spectral data points in
P2 appear to be at the same flux level as the VHE data points. If they are
closely connected, similar behavior might be seen in the VHE emission.
Unfortunately HESS did not conducted any further observations of the galaxy
after year 2009. Hopefully with other VHE facilities at work, possible 
changes of the VHE emission may be detected in the near future, 
which could provide hints for further understanding of 
the \gr\ properties of NGC~253.

\begin{acknowledgements}
This research has made use of the SIMBAD database, operated at CDS,
Strasbourg, France. This research has made use of the MAXI data provided
	by RIKEN, JAXA and the MAXI team. We acknowledge the use of 
	the Fermi Solar Flare Observations facility funded by the Fermi GI 
	program.
	
We thank the referee for detailed suggestions, which helped improve
	this paper.
This research is supported by
the Basic Research Program of Yunnan Province
(No. 202201AS070005), the National Natural Science Foundation of
China (12273033), and the Original
Innovation Program of the Chinese Academy of Sciences (E085021002).
\end{acknowledgements}

\bibliographystyle{aasjournal}
\bibliography{ngc}

\end{document}